\documentclass[12pt]{iopart}
\usepackage{graphicx,epsf}

\def\g{{\gamma}_2^\Lambda}

\def\gw{\mathcal U^\Lambda}
\def\Gl#1{{\mathcal G^\Lambda(#1)}}
\def\Gp#1{{\mathcal S^\Lambda(#1)}}
\def\Se#1{{\Sigma^\Lambda(#1)}}
\def\H#1{{\Theta(|#1|-\Lambda)}}
\def\D#1{{\delta(#1)}}

\def\Gn#1{{\mathcal G^{0}(#1)}}

\def\DS{\displaystyle}
\def\DDL{\frac{d}{d\Lambda}}
\def\gp{\mathcal U_{p}^\Lambda}
\def\ga{\mathcal U_{a}^\Lambda}
\def\gs{\mathcal U_{S}^\Lambda}
\def\gt{\mathcal U_{T}^\Lambda}

\begin{document}
\title{Functional renormalization group approach to zero-dimensional
interacting systems}
\author{R.~Hedden, V.~Meden, Th.~Pruschke, and K.~Sch\"onhammer}
\address{Institut f\"ur Theoretische Physik, Universit\"at G\"ottingen,
  Tammannstr.~1, D-37077 G\"ottingen, Germany}
\ead{meden@theorie.physik.uni-goettingen.de}
\begin{abstract}
We apply the functional renormalization group method to the calculation of 
dynamical properties of zero-dimensional interacting quantum systems. 
As case studies we discuss the anharmonic oscillator and the single 
impurity Anderson model. We truncate the hierarchy of flow equations
such that the results are at least correct up to 
second order perturbation theory in the coupling. For the anharmonic 
oscillator energies and spectra 
obtained within two different functional renormalization group schemes 
are compared to numerically exact results, 
perturbation theory, and the mean field
approximation. Even at large coupling the results obtained using 
the functional renormalization group agree quite well with the
numerical exact solution. The better of the two schemes is used to 
calculate spectra of the single impurity Anderson model, which then 
are compared to the results of perturbation theory and the numerical 
renormalization group. For small to intermediate couplings the
functional renormalization group gives results which are close 
to the ones obtained using the very accurate numerical renormalization 
group method. 
In particulare the low-energy scale (Kondo temperature) extracted
from the functional renormalization group results shows the expected 
behavior.
\end{abstract}
\submitto{\JPC}
\maketitle

\section{Introduction}
\label{sec:intro}
The reliable calculation of physical properties of interacting quantum
mechanical systems presents a formidable task. Typically, one has to 
cope with the interplay of different energy-scales possibly covering 
several orders of magnitude even for simple situations. Approximate 
tools like perturbation theory, but even numerically exact techniques 
can usually handle only a restricted window of energy scales and are 
furthermore limited in their applicability by the approximations 
involved or the computational resources available. In addition due to 
the divergence of certain classes of Feynman diagrams some of the 
interesting many-particle problems cannot be tackled by 
straightforward perturbation theory. 

A general scheme that is designed to handle such multitude of energy
scales and competition of divergences is the renormalization group
\cite{RG_general}. The idea of this approach is to start from high
energy scales, leaving out possible infrared divergences and work 
one's way down to the desired low-energy region in a systematic way. 
The precise definition of ``systematic way'' does in general depend 
on the problem studied.

In particular for interacting quantum mechanical many-particle systems, 
two different schemes attempting a unique, problem independent 
prescription have emerged during the past decade. One is Wegner's  
flow equation technique \cite{wegner}, where a given Hamiltonian is 
diagonalized by continuous unitary transformation. From the final
result one can extract detailed information about the structure of 
the ground state and low-lying excitations. This technique has been 
applied successfully to both fermionic and bosonic problems 
\cite{flow_apps}. However, in general it becomes a rather cumbersome 
task to really calculate physical quantities, especially dynamics 
from correlation functions. Here, one typically has to introduce
further approximations \cite{uhrig,kehrein}, again tightly tailored 
for the problem under investigation.

The second field theoretical approach, which we want to focus on in
the following, is based on a functional representation of the partition 
function of the system under consideration. It has become known 
as functional renormalization group (fRG)
\cite{polchinski,wetterich,morris,salmhoferbuch}. A detailed 
description of the fRG will be given in the next section; here
we make some principle remarks and discuss previous applications.

Different versions of the fRG have been developed over the last few 
years. One either generates an exact infinite hierarchy of coupled
differential equations for the amputated connected $m$-particle Green 
functions of the many-body system \cite{polchinski,salmhoferbuch}, 
or the one-particle 
irreducible $m$-particle vertex functions \cite{wetterich,morris}
respectively. For explicit calculations one has to
truncate the set of equations which is the major approximation 
involved. At what level this truncation is performed to obtain a 
tractable set of equations depends on the complexity of the problem. 
We here exclusively study the one-particle irreducible version of the 
fRG. It has the advantage of including self-energy corrections
already in low truncation orders and being formally easily extendable to 
higher orders \cite{footnote1}. 

Up to now, most applications of the fRG in many-body physics 
concentrate on low-dimensional, interacting fermion systems where 
it provides a possibility to sum  diverging classes of diagrams. The 
homogeneous two-dimensional Hubbard model
\cite{schulz,metzner,honerkamp}, the homogeneous one-dimensional 
Tomonaga-Luttinger model \cite{peter}, and the one-dimensional 
lattice model of spinless 
fermions with nearest neighbor interaction and local impurities
\cite{meden} were investigated. The focus was put on properties of 
the system close to the Fermi surface, for example on the hierarchy of 
interactions to identify possible instabilities 
\cite{schulz,metzner,honerkamp} or on Tomonaga-Luttinger 
liquid exponents \cite{peter,meden}. The frequency
dependence of the vertex functions was mostly neglected
\cite{honerkamp2}. 

In this paper we investigate the frequency dependence of the
self-energy and the effective interaction. For this purpose, we study 
two different zero-dimensional (local) models: 
the quantum anharmonic oscillator and a well-known problem of
solid state physics, the single impurity Anderson model
(SIAM). The former has quite often been used as a ``toy model'' to
investigate the performance of different approximation schemes of
many-particle physics \cite{oscpaper,goetz}. For this problem 
conventional perturbation theory is regular --- although it generates a 
generic example of an asymptotic series \cite{oscpaper} --- and one
expects that compared to perturbation theory the fRG leads to a
better agreement with the exact solution at larger
coupling constants (``renormalization group enhanced perturbation 
theory''). We calculate the ground state energy and the energy of the 
first excited state as well as the spectral function of the propagator. 
Exact results for these observables can numerically be obtained quite 
easily. 

The SIAM has a known hierarchy of energy scales, and presents
a true challenge to any many-body tool due to
the generation of an exponentially small energy scale, the Kondo
scale, leading for example to a sharp resonance in the single-particle 
spectrum. No exact solutions for dynamical quantities 
of the model are known. We present fRG 
results for the one-particle spectral function of the impurity site
and compare them to conventional second order perturbation
theory (in the interaction $U$ of the impurity site) and Wilson's 
numerical renormalization group (NRG). In both models 
the truncated fRG scheme, which is correct at least to second 
order perturbation theory in the interaction, leads to a considerable
improvement compared to plain second order perturbation theory.

The paper is organized as follows. In the next section we present a detailed
discussion of the fRG. The third part contains the application to the
quantum anharmonic oscillator, while in the forth section we 
discuss the fRG for the single impurity Anderson model. A summary
and outlook concludes the paper.

\section{Functional renormalization group}
\label{sec:fRG}
Expressed as a functional integral the grand canonical partition
function of a system of quantum mechanical particles (either fermions
or bosons) interacting via a two-particle potential can be written 
as \cite{negele}
\begin{eqnarray}
\label{partitionfunction}
\frac{{\mathcal Z}}{{\mathcal Z}_0} = \frac{1}{{\mathcal Z}_0} \int
{\mathcal D} \bar \psi \psi \exp{\left\{ (\bar \psi, \left[{\mathcal
       G}^0 \right]^{-1}\psi ) - S_{\rm int}\left(\{ \bar \psi \}, \{
       \psi\} \right) \right\} } \; ,
\end{eqnarray}
with
\begin{eqnarray}
\label{sintallg}
S_{\rm int} \left(\{ \bar \psi \}, \{
  \psi\} \right) = \frac{1}{4}
\sum_{k_1',k_2',k_1,k_2}\bar v_{k_1',k_2',k_1, k_2}
\bar \psi_{k_1'} \bar \psi_{k_2'} \psi_{k_2} \psi_{k_1} \; ,
\end{eqnarray}
and ${\mathcal Z}_0$ being the non-interacting partition function.
Here $\psi$ and $\bar \psi$ denote either Grassmann (fermions) or
complex (bosons) fields. The multi-indices $k_j^{(')}$ stand for
the quantum numbers of the one-particle basis
in which the problem is considered (e.g.\ momenta and spin directions)
and Matsubara frequencies $\omega$. We have introduced the short hand
notation 
\begin{eqnarray*}
\left(\bar \psi,  \left[{\mathcal
       G}^0 \right]^{-1}  \psi \right) = \sum_{k,k'} \bar
    \psi_{k}  \left[{\mathcal
       G}^0 \right]^{-1}_{k,k'} \psi_{k'} \; ,
\end{eqnarray*}
with the propagator ${\mathcal G}^0$ of the related non-interacting
problem given as a matrix. The anti-symmetrized
(fermions) or symmetrized (bosons) matrix elements of the two-particle 
interaction are denoted by $\bar v_{k_1',k_2',k_1, k_2}$. They
contain the energy conserving factor 
$\delta_{\omega+\omega',\nu+\nu'}$ and the factor 
$1/\beta$, with $\beta=1/T$ being the inverse temperature. We consider
units such that $\hbar=1$ and $k_B=1$. The generating functional of
the $m$-particle Green function is given by
\begin{eqnarray}
\label{genfunct}
{\mathcal W } \left(\{\bar \eta \}, \{ \eta\} \right) & = &
\frac{1}{{\mathcal Z}_0} 
\int {\mathcal D} \bar \psi \psi  \exp \left\{ 
 \left( \bar \psi, \left[ {\mathcal G}^{0}\right]^{-1}
   \psi \right) - S_{\rm int}(\{\bar \psi \}, \{ \psi \})   \nonumber
\right. \\
&& \left. 
  - \left( \bar \psi, \eta\right) - \left(\bar \eta,  \psi  
   \right) \right\} \; ,
\end{eqnarray}
with $\left( \bar \psi, \eta \right) = \sum_{k} \bar \psi_{k}
\eta_{k}$ and the external source fields $\eta$ and $\bar \eta$. 
From this the generating functional of
the connected $m$-particle Green function follows as 
\begin{eqnarray}
\label{wcdef}
{\mathcal W}^c\left(\{\bar \eta \}, \{ \eta\} \right) = 
\ln{ \left[ {\mathcal W}
\left(\{\bar \eta \}, \{ \eta\} \right) \right]} \; .
\end{eqnarray}
The (connected) $m$-particle Green function $G_m^{(c)}$ can be
obtained by taking functional derivatives 
\begin{eqnarray}
\label{gmdef}
\hspace{-1.5cm} G_m^{(c)}\left(k_1', \ldots, k_m'; k_1, \ldots, k_m \right) = 
\left. \frac{\delta^m}{\delta \bar \eta_{k_1'} 
\ldots \delta \bar \eta_{k_m'}} 
\frac{\delta^m}{\delta  \eta_{k_m} \ldots  \delta  \eta_{k_1}} {\mathcal
  W}^{(c)}\left(\{\bar \eta \}, \{ \eta\} \right) \right|_{\eta = 0= \bar
\eta} \; .
\end{eqnarray}
By a Legendre transformation 
\begin{eqnarray}
\label{gammadef}
\Gamma \left(\{\bar \phi \}, \{ \phi \} \right) = -  {\mathcal
  W}^c\left(\{\bar \eta \}, \{ \eta\} \right) - \left(\bar \phi, \eta
  \right) -\left(\bar \eta, \phi \right) + \left(\bar \phi, \left[{\mathcal
  G}^{0}\right]^{-1} \phi\right)  \; ,
\end{eqnarray} 
the generating functional of the one-particle irreducible vertex 
functions $\gamma_m$, with external source fields $\phi$ and $\bar
\phi$ and
\begin{eqnarray}
\label{vertexfundef}
\hspace{-1.5cm}\gamma_m\left(k_1', \ldots, k_m'; k_1, \ldots, k_m \right) & = &
\left. \frac{\delta^m}{\delta \bar \phi_{k_1'} \ldots \delta \bar \phi_{k_m'}}
\frac{\delta^m}{\delta  \phi_{k_m} \ldots \delta  \phi_{k_1}} \Gamma
\left(\{\bar \phi\}, \{ \phi\} \right) \right|_{\phi = 0= \bar
\phi} \; ,
\end{eqnarray}
is obtained. Note that in contrast to the usual
definition \cite{negele} of $\Gamma $ 
we have added a term $\left(\bar \phi, \left[{\mathcal G}^{0}\right]^{-1}
  \phi\right)$ in Eq.\ (\ref{gammadef}) for convenience (see below). 
The relation between the
$G_m^{(c)}$ and $\gamma_m$ can be found in text books \cite{negele}.
The 0-particle vertex $\gamma_0$ provides the interacting part of the
grand canonical potential $\Omega$
\begin{eqnarray*}
\Omega=- T \ln {\mathcal Z} = T \gamma_0 -  T \ln {\mathcal Z}_0 \; .
\end{eqnarray*}
For the 1-particle Green function we obtain
\begin{eqnarray*}
G_1(k';k) = G_1^c(k';k) = - \zeta {\mathcal G}_{k',k}
=  \left[  \gamma_1 - \zeta \left[ {\mathcal G}^0\right]^{-1}
  \right]^{-1}_{k',k} \; ,
\end{eqnarray*}
where
\begin{eqnarray*}
{\mathcal G}_{k',k}  =
\left[   \left[ {\mathcal G}^0\right]^{-1}
 - \Sigma   \right]^{-1}_{k',k} \; ,
\end{eqnarray*}
with the self-energy $\Sigma$, 
and $\zeta=-1$ for fermions or $\zeta=1$ for bosons, respectively. 
This implies the relation $\Sigma = \zeta \gamma_1$. The 2-particle 
vertex $\gamma_2$ is usually referred to as the effective interaction. 
For $m \geq 3$, the $m$-particle interaction $\gamma_m$ has
diagrammatic contributions starting at 
$m$-th order in the two-particle interaction. 

In Eqs.\ (\ref{partitionfunction}) and (\ref{genfunct}) we 
replace the non-interacting propagator by a propagator ${\mathcal
  G}^{0,\Lambda}$  depending on a cutoff $\Lambda$
and ${\mathcal Z}_0$
by  ${\mathcal Z}_0^{\Lambda}$ determined using 
${\mathcal G}^{0,\Lambda}$. 
The boundary conditions for the cutoff 
$\Lambda \in [\Lambda_0,0]$ are taken as
\begin{eqnarray}
\label{demands}
{\mathcal G}^{0,\Lambda_0} = 0  \;\;\; , \;\;\;
 {\mathcal G}^{0,\Lambda=0} = {\mathcal G}^{0} \; , 
\end{eqnarray} 
i.e.\ at the starting point $\Lambda = \Lambda_0$ no degrees of
freedom are ``turned on'' while at $\Lambda=0$ the cutoff
independent problem is recovered. 
In our applications we use a sharp Matsubara frequency cutoff 
with 
\begin{eqnarray}
\label{matsubaracutoff}
{\mathcal G}^{0,\Lambda} = \Theta\left(|\omega|-\Lambda\right)
{\mathcal G}^{0}  
\end{eqnarray} 
and consider $\Lambda_0 \to \infty$ \cite{meden}. Through
${\mathcal G}^{0,\Lambda}$ the quantities defined in Eqs.\
(\ref{partitionfunction}) to
(\ref{vertexfundef}) acquire a $\Lambda$-dependence. 
One now derives a functional differential equation for 
$\Gamma^{\Lambda}$. From this, by expanding in powers of the 
external sources, a set of coupled differential equations for the 
$\gamma_m^\Lambda$ is obtained. 

As a first step we 
differentiate ${\mathcal W}^{c,\Lambda}$ 
with respect to $\Lambda$, which after straightforward algebra leads to 
\begin{eqnarray}
\label{flusswc}
\hspace{-1cm}\frac{d}{d \Lambda} {\mathcal W}^{c,\Lambda} & = &  
\zeta \mbox{ Tr}\, \left( 
{\mathcal Q}^{\Lambda} {\mathcal G}^{0,\Lambda}
 \right) + 
\mbox{Tr} \, 
\left({\mathcal Q}^{\Lambda} 
 \frac{\delta^2
   {\mathcal W}^{c,\Lambda} }{ \delta \bar \eta \delta
      \eta}  \right) + \zeta \left(
   \frac{\delta {\mathcal W}^{c,\Lambda} }{\delta \eta},
   {\mathcal Q}^{\Lambda}  \frac{\delta {\mathcal
       W}^{c,\Lambda} }{\delta \bar \eta} \right) \; ,
\end{eqnarray}
with 
\begin{eqnarray}
\label{Qlambdadef}
 {\mathcal Q}^{\Lambda} = \frac{d}{d \Lambda} \left[ {\mathcal
 G}^{0,\Lambda}\right]^{-1} \; .
\end{eqnarray} 
Considering $\phi$ and $\bar \phi$ as the fundamental variables we
obtain from Eq.\ (\ref{gammadef}) 
\begin{eqnarray*}
\hspace{-2.3cm}\frac{d}{d \Lambda} \Gamma^{\Lambda}\left(\{\bar \phi \}, \{ \phi \} \right) = 
- \frac{d}{d \Lambda} {\mathcal W}^{c,\Lambda} \left(\{\bar
  \eta^{\Lambda} \}, \{ \eta^{\Lambda} \} \right) 
- \left(\bar \phi, \frac{d}{d \Lambda} \eta^{\Lambda}
  \right) -\left( \frac{d}{d \Lambda} {\bar \eta}^{\Lambda} , \phi \right) 
+ \left(\bar \phi,  {\mathcal Q}^{\Lambda}  \phi\right) \; .
\end{eqnarray*}
Applying the chain rule and using Eq.\ (\ref{flusswc}) this leads to 
\begin{eqnarray*}
\frac{d}{d \Lambda} \Gamma^{\Lambda} = - \zeta \mbox{ Tr}\, \left(
 {\mathcal Q}^{\Lambda} 
{\mathcal
     G}^{0,\Lambda}  \right) 
- \mbox{Tr} \, \left(  {\mathcal Q}^{\Lambda} 
 \frac{\delta^2
   {\mathcal W}^{c,\Lambda} }{ \delta \bar \eta^{\Lambda} \delta
      \eta^{\Lambda}}  \right) \; .
\end{eqnarray*}
The last term in Eq.\  (\ref{gammadef}) is exactly cancelled, which a
posteriori justifies its inclusion.
Using the well known relation \cite{negele} between the second
functional derivatives of $\Gamma$ and ${\mathcal W}^{c}$ 
we obtain the functional differential equation 
\begin{eqnarray}
\label{gammafluss}
\frac{d}{d \Lambda} \Gamma^{\Lambda} = - \zeta 
\mbox{ Tr}\, \left( {\mathcal Q}^{\Lambda}  {\mathcal
     G}^{0,\Lambda} 
 \right) - \mbox{Tr} \, \left({\mathcal Q}^{\Lambda}   
 {\mathcal V}_{\bar \phi, \phi}^{1,1}(\Gamma^{\Lambda}, 
{\mathcal  G}^{0,\Lambda})  \right) \; ,
\end{eqnarray}
where $ {\mathcal V}_{\bar \phi, \phi}^{1,1}$ stand for the upper left
block of the matrix
\begin{eqnarray}
\label{invers}
{\mathcal V}_{\bar \phi,\phi}(\Gamma^{\Lambda}, {\mathcal G}^{\Lambda}) = 
\left( \begin{array}{cc} 
\frac{\delta^2 \Gamma^{\Lambda}}{ \delta \bar \phi \delta \phi}
   - \zeta \left[ {\mathcal G}^{0,\Lambda} \right]^{-1}
&  \frac{\delta^2 \Gamma^{\Lambda}}{\delta \bar \phi \delta  \bar \phi} \\
\frac{\delta^2 \Gamma^{\Lambda}}{\delta  \phi \delta  \phi} 
&  \frac{\delta^2 \Gamma^{\Lambda}}{\delta \phi \delta \bar \phi} - 
\left[ \left[{\mathcal G}^{0,\Lambda} \right]^{-1}\right]^t 
\end{array} \right)^{-1} 
\end{eqnarray}
and the upper index $t$ denotes the transposed matrix.
To obtain differential equations for the $\gamma_m^{\Lambda}$ which
include self-energy corrections we express $ {\mathcal V}_{\bar \phi,
  \phi}$  in terms of
${\mathcal G}^{\Lambda}$ instead of ${\mathcal G}^{\Lambda,0}$.
This is achieved by defining 
\begin{eqnarray*}
{\mathcal U}_{\bar \phi,   \phi} = \frac{\delta^2 \Gamma^{\Lambda}}{\delta \bar
  \phi \delta \phi} - \gamma_1^{\Lambda}  
\end{eqnarray*}
and using
\begin{eqnarray}
\label{GG}
{\mathcal G}^{\Lambda} = \left[ \left[ {\mathcal G}^{0,\Lambda}
  \right]^{-1} -  \zeta \gamma_1^{\Lambda}  \right]^{-1} 
\end{eqnarray}
which leads to
\begin{eqnarray}
\label{gammaflussalt}
\frac{d}{d \Lambda} \Gamma^{\Lambda} = - \zeta \mbox{ Tr}\, \left(
  {\mathcal Q}^{\Lambda}  {\mathcal
     G}^{0,\Lambda} 
 \right) + \zeta \mbox{Tr} \, \left({\mathcal G}^{\Lambda} 
   {\mathcal Q}^{\Lambda} 
 \tilde{\mathcal V}_{\bar \phi, \phi}^{1,1}(\Gamma^{\Lambda},{\mathcal
   G}^{\Lambda} )  \right)
\; , 
\end{eqnarray} 
with 
\begin{eqnarray}
\label{Vtildedef}
\tilde {\mathcal V}_{\bar \phi,\phi}\left(\Gamma^{\Lambda}
,{\mathcal  G}^{\Lambda}  \right)  = \left[ {\bf 1} - 
\left( \begin{array}{cc} 
\zeta {\mathcal G}^{\Lambda} 
&  
0\\
0
&  
\left[ {\mathcal G}^{\Lambda} \right]^t 
\end{array} \right)
\left( \begin{array}{cc} 
{\mathcal U}_{\bar \phi,   \phi}
&  \frac{\delta^2 \Gamma^{\Lambda}}{\delta \bar \phi \delta  \bar \phi} \\
\frac{\delta^2 \Gamma^{\Lambda}}{\delta  \phi \delta  \phi} 
&  \zeta  {\mathcal U}^{t}_{\bar \phi,   \phi}
\end{array} \right) \right]^{-1} \; .
\end{eqnarray} 
For later applications it is important to note that ${\mathcal U}_{\bar \phi,
  \phi}$ as well as $ \frac{\delta^2
  \Gamma^{\Lambda}}{\delta \bar \phi \delta \bar \phi}$ and $\frac{\delta^2
  \Gamma^{\Lambda}}{\delta  \phi \delta  \phi} $ are at least
quadratic in the external sources. 
The initial condition for the exact functional differential equation 
(\ref{gammaflussalt}) can either be obtained by lengthy but 
straightforward algebra not presented here, or 
by the following simple 
argument: at $\Lambda= \Lambda_0$, ${\mathcal G}^{0,\Lambda_0}=0$ (no
degrees of freedom are ``turned on'') and in a perturbative expansion
of the $\gamma_m^{\Lambda_0}$ the only term which does not vanish is the
bare 2-particle vertex. We thus find
 \begin{eqnarray}
\label{anfanggamma}
\Gamma^{\Lambda_0} \left( \{ \bar \phi \} , \left\{  \phi
  \right\} \right) = S_{\rm int}  \left(\{\bar
   \phi \}, \{ \phi \} \right) \; .
\end{eqnarray} 

\begin{figure}[htb]
\centerline{\includegraphics[width=0.28\textwidth,clip]{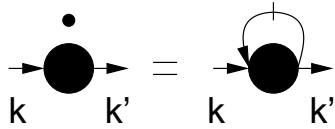}}
\caption[]{Diagrammatic form of the flow equation for
  $\gamma_1^{\Lambda} = \zeta \Sigma^{\Lambda}$. The slashed line
  stands for the single scale propagator ${\mathcal S}^{\Lambda}$. 
\label{fig1}}
\end{figure}

\begin{figure}[htb]
\centerline{\includegraphics[width=0.6\textwidth,clip]{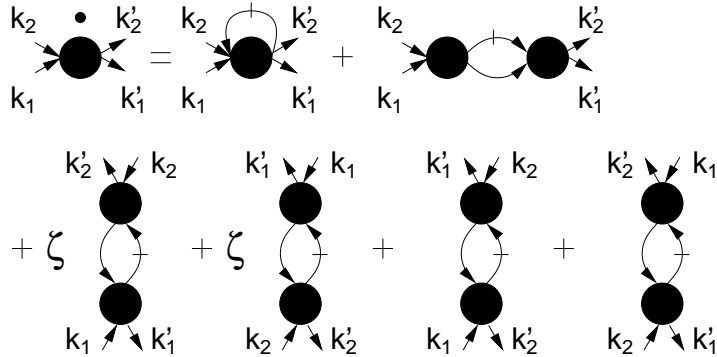}}
\caption[]{Diagrammatic form of the flow equation for
  $\gamma_2^{\Lambda}$. The slashed line
  stands for the single scale propagator ${\mathcal S}^{\Lambda}$ the
  unslashed line for ${\mathcal G}^{\Lambda}$. 
\label{fig2}}
\end{figure}

An exact infinite hierarchy of flow equations 
for the $\gamma_m^{\Lambda}$ can be obtained by 
expanding Eq.\ (\ref{Vtildedef}) in a geometric series and
$\Gamma^{\Lambda}$ in the external sources
\begin{eqnarray*}
\Gamma^{\Lambda}\left(\{\bar \phi \}, \{ \phi \} \right) =
\sum_{m=0}^{\infty} \frac{\zeta^m}{(m !)^2}  \sum_{k_1', \ldots, k_m'} 
\sum_{k_1, \ldots, k_m } && \gamma_m^{\Lambda}\left(k_1',
  \ldots, k_m'; k_1, \ldots, k_m \right) \nonumber \\ 
&& \times \bar \phi_{k_1'} \ldots \bar
\phi_{k_m'} \phi_{k_m} \ldots  \phi_{k_1} \; .
\end{eqnarray*}
The equation for
$\gamma_0^{\Lambda}$ reads 
\begin{eqnarray}
\label{gamma0fl}
\frac{d}{d \Lambda}\gamma_0^{\Lambda} = - \zeta \mbox{ Tr}\, \left( 
   {\mathcal Q}^{\Lambda} {\mathcal
     G}^{0,\Lambda}  \right) +
\zeta \mbox{ Tr}\, \left(  {\mathcal Q}^{\Lambda} {\mathcal
     G}^{\Lambda}  \right) \; .
\end{eqnarray}
Via ${\mathcal G}^{\Lambda}$ the derivative of $\gamma_0^{\Lambda}$
couples to the one-particle self-energy. For the flow of the 
self-energy we obtain
\begin{eqnarray}
\label{gamma1fl}
\frac{d}{d \Lambda} \gamma_1^{\Lambda}(k';k) = 
\zeta \frac{d}{d \Lambda} \Sigma^{\Lambda}_{k',k}
= \mbox{ Tr}\, \left( {\mathcal S}^{\Lambda} 
 \gamma_2^{\Lambda}(k', \ldots; 
k, \ldots) \right) \; , 
\end{eqnarray}  
with the so-called single scale propagator
\begin{eqnarray}
\label{Sdef}
{\mathcal S}^{\Lambda}  =
{\mathcal G}^{\Lambda} {\mathcal Q}^{\Lambda}   
{\mathcal G}^{\Lambda}  \; .
\end{eqnarray}  
Here $\gamma_2^{\Lambda}(k', \ldots; 
k, \ldots)$ is a matrix in the variables not explicitly
written, i.e.\ $\left[\gamma_2^{\Lambda}(k', \ldots ;k,\ldots
  )\right]_{q',q} = \gamma_2^{\Lambda}(k',q'; k,q) $. Diagrammatically
Eq.\ (\ref{gamma1fl}) is shown in Fig.\ \ref{fig1}. The derivative 
of $ \gamma_1^{\Lambda}$ is determined by $ \gamma_1^{\Lambda}$ and the
2-particle vertex $ \gamma_2^{\Lambda}$. Thus an equation for $
\gamma_2^{\Lambda}$ is required 
\begin{eqnarray}
\label{gamma2fl}
\hspace{-0.8cm} \frac{d}{d \Lambda} \gamma_2^{\Lambda}(k_1', k_2'; k_1, k_2) 
& = & \mbox{ Tr}\, \left( {\mathcal S}^{\Lambda}
\gamma_3^{\Lambda}(k_1',k_2', \ldots; k_1,k_2,\ldots) 
\right) 
\nonumber \\ 
&&  + \zeta \mbox{ Tr}\, \Big(  {\mathcal S}^{\Lambda}
  \gamma_2^{\Lambda}(\ldots , \ldots ; k_1, k_2 )
\left[{\mathcal G}^{\Lambda}\right]^{t}   
\gamma_2^{\Lambda}(k_1', k_2' ; \ldots
,\ldots ) \Big)
\nonumber \\ 
&&  + \zeta \mbox{ Tr}\, \Big( {\mathcal S}^{\Lambda} 
 \gamma_2^{\Lambda}(k_1', \ldots ;
k_1,\ldots ) 
{\mathcal G}^{\Lambda}   \gamma_2^{\Lambda}(k_2', \ldots ;
k_2,\ldots )  \nonumber \\
&&   + \zeta \left[ k_1' \leftrightarrow k_2' \right] 
 + \zeta \left[ k_1 \leftrightarrow k_2 \right] + 
 \left[ k_1' \leftrightarrow k_2' ,  k_1 \leftrightarrow k_2 
\right]  \Big)
 \; .
\end{eqnarray}
The corresponding diagrammatic representation is shown in Fig.\
\ref{fig2}. The right hand side (rhs) of Eq.\ (\ref{gamma2fl}) contains 
$\gamma_1^{\Lambda}$, $\gamma_2^{\Lambda}$, and the 3-particle 
vertex $\gamma_3^{\Lambda}$. For $m \geq 1$ the equation 
for $\frac{d}{d \Lambda} 
\gamma_m^{\Lambda}$ contains $\gamma_{m'}^{\Lambda}$ 
with $m'=1,2,\ldots,m, m+1$. The initial condition  
for the $\gamma_m^{\Lambda_0}$ can be obtained from Eq.\
(\ref{anfanggamma}) and is given by
\begin{eqnarray}
\gamma_2^{\Lambda_0}(k_1',k_2';k_1,k_2) =  \bar v_{k_1',k_2',k_1, k_2}
\;\;\; , \;\;\; \gamma_m^{\Lambda_0} = 0 \;\; \mbox{for} \;\;  m \neq 2 \; . 
\label{gammamanfang} 
\end{eqnarray} 
We here refrain from explicitly presenting equations for 
$\frac{d}{d \Lambda} \gamma_m^{\Lambda}$ with $m \geq 3$ since later
on the set of differential equations is truncated by setting 
$\gamma_3^{\Lambda} = \gamma_3^{\Lambda_0} =0$, which implies that 
vertices with $m \geq 3$ do not contribute.

Following the above systematics, a truncation scheme emerges quite
naturally. If, for $m_c \geq 2$, the vertex $\gamma_{m_c+1}^{\Lambda}$ 
on the rhs of the coupled flow equations is replaced by its initial 
condition $\gamma_{m_c+1}^{\Lambda_0}=0$, a closed set
of equations for $\gamma_{m}^{\Lambda}$ with $m \leq m_c$
follows. This set of differential equations can then be integrated
over $\Lambda$ starting at $\Lambda=\Lambda_0$ down to
$\Lambda=0$ providing approximate expressions for the $\gamma_{m}$ of
the original (cutoff free) problem with $m \leq m_c$. 
Expanding $\gamma_{m}^{\Lambda}$ in terms of the bare
interaction, conventional perturbation theory for the grand canonical
potential, the self-energy, the effective interaction and higher order
vertex functions can be recovered from an iterative treatment of the 
flow equations. It is easy
to show that the vertex functions obtained from the truncated
equations are at least correct up to order $m_c$ in the bare
interaction. 

To obtain a manageable set of equations in applications
of the fRG to one- and two-dimensional quantum many-body problems
\cite{schulz,metzner,honerkamp,peter,meden}, further approximations in
addition to the above 
truncation scheme were necessary. In the following two sections we
will avoid such additional approximations and solve the truncated fRG 
equations to order $m_c=2$ for two zero-dimensional (local) interacting
many-particle problems: the quantum harmonic oscillator with a quartic
perturbation and the SIAM.  

Very recently a modified version of the flow equation for 
$\gamma_2^{\Lambda}$ was suggested. As we will consider the truncation
order $m_c=2$ we here only describe this modified scheme for this
order. Guided by the idea of an improved fulfillment (compared to the
scheme described above) of a certain Ward identity 
Katanin replaced the combined propagator ${\mathcal
  S}^{\Lambda}$  in the last five terms of Eq.\ (\ref{gamma2fl}) (the
first term does not contribute to order $m_c=2$) by  
$-d {\mathcal G}^{\Lambda} / d\Lambda$ \cite{katanin}. Using 
\begin{eqnarray*}
\frac{d {\mathcal G}^{\Lambda}}{d \Lambda} = - {\mathcal S}^{\Lambda}
+ {\mathcal G}^{\Lambda} \frac{d \Sigma^{\Lambda}}{d \Lambda}
{\mathcal G}^{\Lambda} 
\end{eqnarray*}
obtained from Eq.\ (\ref{GG}) it is obvious that the terms added 
to Eq.\ (\ref{gamma2fl}) are at least of third order
in the bare interaction. Using this modification in the order $m_c=2$ equations for
$\gamma_1^{\Lambda}$ and $\gamma_2^{\Lambda}$ 
leads  to the exact solution for certain exactly solvable models (reduced BCS
model, interacting fermions with forward scattering only).  
This suggests that the replacement possibly improves
the results of the truncated fRG also for other models. In
the next section we show that this is indeed the case for the
harmonic oscillator with a quartic perturbation. The same holds for
the SIAM as discussed in Sect.\ 4. 

\section{Application to the anharmonic oscillator}
\label{sec:ao}
In appropriate units the Hamiltonian of the harmonic oscillator 
with a quartic perturbation is given by
\begin{equation}
\label{anhaos}
H=\frac{1}{2} x^2 + \frac{1}{2} p^2 + \frac{g}{4!} x^4 \; ,
\end{equation}
with the position operator $x$, the momentum operator $p$, 
and the coupling constant $g$. 
We here focus on $T=0$ and are interested in low-lying eigenenergies 
$E_n$ as well as the (imaginary) time-ordered propagator 
\begin{equation*}
{\mathcal G}(\tau) = \left< E_0 \right| {\mathcal T} 
\left[x(\tau) x(0) \right] \left| E_0 \right> \; , 
\end{equation*}
with $\left| E_n \right>$ being the eigenstates of the Hamiltonian in 
Eq.\ (\ref{anhaos}). The Fourier transform of the propagator can be written as
\begin{eqnarray*}
{\mathcal G} (i \omega) = 
\int_{-\infty}^{\infty} d \tau \, e^{i \omega \tau} {\mathcal G}(\tau) 
= \frac{1}{\left[ {\mathcal G}^0 (i \omega) 
\right]^{-1}- \Sigma(i\omega)} \; ,
\end{eqnarray*}
where we have introduced the self-energy $\Sigma$
and the non-interacting propagator 
\begin{eqnarray*}
{\mathcal G}^0 (i \omega)  =  
\frac{1}{\omega^2 + 1} \; . 
\end{eqnarray*}  
In contrast to the more general notation used in the last section,
propagators and the self-energy only depend on a single frequency
and do no longer contain the energy conserving $\delta$-function ($T=0$) here.
The propagator has the Lehmann representation
\begin{eqnarray}
\label{lehmann}
\hspace{-1.0cm} {\mathcal G} (i \omega)  =  
\frac{1}{2} \sum_{n=1}^{\infty} \left[
  \frac{1}{i \omega + (E_n-E_0)} -  \frac{1}{i \omega - (E_n-E_0)}
\right] \left| \left< E_0 \right| (a +a^{\dag}) \left| E_n \right>
\right|^2 \; ,
\end{eqnarray}
where $a$, $a^\dag$ denote the usual raising and lowering operators.  
The spectral weights 
\begin{eqnarray*}
s_n = \left| \left< E_0 \right| (a +a^{\dag}) \left| 
E_n \right> \right|^2
\end{eqnarray*} 
 and energies $E_n$ fulfill the f-sum rule 
\begin{eqnarray}
\label{sumrule}
1 =  \sum_{n=1}^{\infty} (E_n-E_0) \; s_n \; .
\end{eqnarray}
It turns out that for coupling constants $g \leq 50$ considered here
the sums in Eqs.\ (\ref{lehmann}) and  (\ref{sumrule}) are dominated 
by the first term. For this reason only the first few 
eigenstates and eigenenergies are required to obtain accurate (``numerically
exact'') results for ${\mathcal G} (i \omega)$. These can quite easily 
be obtained by expressing  $H$ in the basis of eigenstates 
$\left| n \right>$ of the unperturbed harmonic oscillator and 
numerically diagonalizing the upper left corner of the
(infinite) matrix $\left< n \right| H \left| n' \right>$ with 
$n,n' \leq n_c$ and a sufficiently large $n_c$. For $g \leq 50$,
$n_c=100$ turns out to be large enough to fulfill the sum rule 
Eq.\ (\ref{sumrule})  to very high precision.

Second order perturbation theory 
for the $g$-dependent part of the ground state energy yields
\begin{eqnarray}
\label{energy2}
e_0^{(2)} = E_0^{(2)} - E_0^0= 
\frac{1}{32} \; g - \frac{7}{1536} \; g^2
\end{eqnarray}
and for the self-energy one obtains
\begin{eqnarray}
\label{selfenergy2}
 \Sigma^{(2)}(i\omega) = 
- \frac{1}{4} \; g + \frac{1}{32} \; g^2 + \frac{1}{8} \; g^2 \frac{1}{\omega^2
  +9} \; .
\end{eqnarray}
Within the fRG approximate expressions for 
\begin{eqnarray*}
E_{n,0}= E_n - E_0
\end{eqnarray*}
and $s_n$ 
can be obtained from the
poles and residues of the propagator ${\mathcal G} (i
\omega)$. Furthermore, since  Eqs.\ (\ref{lehmann}) and
(\ref{sumrule})  are dominated by the first terms we only consider 
$E_{1,0}$ and $s_1$ \cite{footnote4}. Second order approximations for these 
quantities are given by the smallest pole of 
${\mathcal G}^{(2)}(i \omega) = \left[ \omega^2+1 -
\Sigma^{(2)}(i\omega) \right]^{-1}$
and the related residue. It is important
to note that this approximation $E_{1,0}^{(2)}$ agrees with 
$E_1^{(2)}-E_0^{(2)}$, where $E_1^{(2)}$ is determined directly from
Rayleigh-Schr\"odinger perturbation theory, only up to second order in
$g$, but is closer to the exact $E_{1,0}$.  

Within mean field theory, $e_0^{\rm MF}$ and
a frequency independent $\Sigma^{\rm MF}$ are given by
\begin{eqnarray*}
e_0^{\rm MF} = \frac{1}{2} \sqrt{1+\frac{g}{2} \left< X^2
  \right>_{\rm MF}} \;  - \frac{g}{8} \left< X^2 \right>_{\rm MF}^2 -
\frac{1}{2} \; \;
\; , \; \; \; \Sigma^{\rm MF} = -\frac{g}{2} \left< X^2
\right>_{\rm MF} 
\end{eqnarray*}
and $\left< X^2 \right>_{\rm MF}$ is the solution of the self-consistency
equation 
\begin{eqnarray*}
\left< X^2 \right>_{\rm MF} = \frac{1}{2} \; \frac{1}{\sqrt{1+g \left<
      X^2 \right>_{\rm MF} /2}} \; . 
\end{eqnarray*}
From the mean field propagator one obtains
\begin{eqnarray*}
E_{1,0}^{\rm MF} = \sqrt{1 - \Sigma^{\rm MF}} \; \;
\; , \; \; \; s_1^{\rm MF} = \frac{1}{2} \;  \frac{1}{ \sqrt{1 -
    \Sigma^{\rm MF}}} \; .
\end{eqnarray*}

The functional integral representation of the grand canonical
partition function of the Hamiltonian Eq.\ (\ref{anhaos}) reads
\begin{eqnarray}
\label{partitionfunctionao}
\frac{{\mathcal Z}}{{\mathcal Z}_0} = \frac{1}{{\mathcal Z}_0} \int
{\mathcal D} \, \bar x \, x \exp{\left\{ (\bar x, \left[{\mathcal
       G}^0 \right]^{-1} x )/2 - S_{\rm int}\left(\{ \bar x \}, \{
       x \} \right) \right\} } \; ,
\end{eqnarray}
with 
\begin{eqnarray}
\label{sintao}
S_{\rm int} \left(\{ \bar x \}, \{
  x \} \right) = \frac{g}{\beta \, 4!}
\sum_{n_1,\ldots , n_4} \delta_{n_1+n_2+n_3+n_4,0}  \;  
x(i \omega_1) x(i \omega_2) x(i \omega_3) x(i \omega_4)  \; ,
\end{eqnarray}
bosonic Matsubara frequencies $\omega_j=2 \pi \, n_j/\beta$, 
and complex fields 
$\bar x( i \omega) = x(-i \omega) $. As outlined 
in the last section and using a frequency cutoff Eq.\
(\ref{matsubaracutoff}) flow equations for the $\gamma_m^{\Lambda}$ 
can be obtained. Here we focus on the equations in truncation order 
$m_c=2$. For $T \to 0$ and after introducing 
\begin{eqnarray*}
e_0^{\Lambda} = \lim_{T \to 0} T \gamma_0^{\Lambda}
\end{eqnarray*}
we find 
\begin{eqnarray}
\label{gamma0ao}
\frac{d}{d \Lambda} e_0^{\Lambda}
= - \frac{1}{2 \pi} \ln \left[ 1-  {\mathcal
       G}^0(i \Lambda) \; \Sigma^{\Lambda}(\Lambda) \right] \; ,
\end{eqnarray}
with the initial condition $e_0^{\Lambda=\infty}=0$.
At the end of the flow, $e_0^{\Lambda=0}$ directly provides the fRG
approximation $e_0^{\rm fRG}$ for the $g$-dependent part 
of the ground state energy. 
The flow equation for the self-energy follows as 
\begin{eqnarray}
\label{gamma1ao}
\frac{d }{d \Lambda} \Sigma^{\Lambda} (i\omega)= 
 \frac{1}{2 \pi} \; \frac{1}{\Lambda^2 + 1 -  \Sigma^{\Lambda}(
   i \omega)} 
\; g^{\Lambda}(i\omega,-i\omega,i\Lambda, -i\Lambda)
  \; ,
\end{eqnarray}
with the initial condition $\Sigma^{\Lambda=\infty}=0 $ and the fRG 
approximation for the self-energy 
$\Sigma^{\rm fRG}(i\omega)=\Sigma^{\Lambda=0}(i\omega)$. 
Here $g^{\Lambda}$ denotes the totally symmetric  
2-particle vertex which, in contrast to the vertex 
$\gamma_2^{\Lambda}$ introduced in the last section, does not 
contain an energy conserving $\delta$-function and factors of
$\beta$. It depends on only three frequencies, but 
the fourth will nevertheless always be included in the following.
To derive Eqs.\ (\ref{gamma0ao}) and (\ref{gamma1ao})  one
has to deal with products of 
delta functions $\delta(|\omega| - \Lambda)$ and terms involving 
step functions $\Theta(|\omega| - \Lambda)$. These seemingly  
ambiguous expressions are well defined and unique if the sharp 
cutoff is implemented as a limit of increasingly sharp broadened 
cutoff functions $\Theta_{\epsilon}$, with the broadening parameter 
$\epsilon$ tending to zero. The expressions can then be conveniently 
evaluated by using the following relation \cite{morris}, valid 
for arbitrary continuous functions $f$:
\begin{equation}
\label{morristrick}
 \delta_{\epsilon}(x-\Lambda) \, f[\Theta_{\epsilon}(x-\Lambda)] \to
 \delta(x-\Lambda) \int_0^1 f(t) \, dt \; ,
\end{equation}
where $\delta_{\epsilon} = - d \Theta_{\epsilon}/ d \epsilon$. 

For $g^{\Lambda}$ the flow equation reads
\begin{eqnarray}
\label{gamma2ao}
&& \hspace{-2.0cm}\frac{d }{d \Lambda} g^{\Lambda} (i\omega_1, i
\omega_2,i\omega_3, -i \omega_1-i \omega_2-i \omega_3 )  = 
 \frac{1}{2 \pi} \int_{-\infty}^{\infty}  d \, \nu \left[  
{\mathcal P}(i \nu , i \nu - i\omega_1 - i \omega_2)
\right. \nonumber \\&& \hspace{-1.5cm} \times 
g^{\Lambda} 
(i\omega_1, i\omega_2,- i \nu, i \nu - i\omega_1 - i \omega_2) 
 g^{\Lambda} (i\omega_3, -i 
\omega_1-i \omega_2-i \omega_3,
-i\nu +i\omega_1 + i \omega_2, i \nu ) \nonumber \\
&& \hspace{-1.5cm}  \left. +\left(  \omega_2  \leftrightarrow \omega_3 \right)
+\left(  \omega_2  \leftrightarrow  -\omega_1- \omega_2- \omega_3 
\right)
\right]
\end{eqnarray}
where ${\mathcal P}(i \nu, i \nu')$ stands for two different
products of propagators. In the conventional fRG scheme it is given by  
\begin{equation}
\label{Pconventional}
{\mathcal P}_{\rm con}(i \nu, i \nu') = {\mathcal S}^{\Lambda}(i \nu) 
\; {\mathcal G}^{\Lambda}(i \nu')   
\end{equation}
while in the modified scheme \cite{katanin} one obtains 
\begin{equation}
\label{Pmodified}
{\mathcal P}_{\rm mod}(i \nu, i \nu') = - \frac{d {\mathcal G}^{\Lambda}(i
  \nu)}{d \Lambda} \;   
{\mathcal G}^{\Lambda}(i \nu') \; .   
\end{equation}
To explicitly evaluate ${\mathcal P}(i \nu, i \nu') $ Eq.\
(\ref{morristrick}) has to be used, where special care has to be taken for 
the case $\nu=\nu'$. In the conventional scheme 
${\mathcal P}_{\rm con}(i \nu, i \nu')$ contains a factor 
$\delta(|\nu|-\Lambda)$ and the integral over $\nu$ in 
Eq.\ (\ref{gamma2ao}) can be performed analytically.

\begin{figure}[htb]
\centerline{\includegraphics[width=0.8\textwidth,clip]{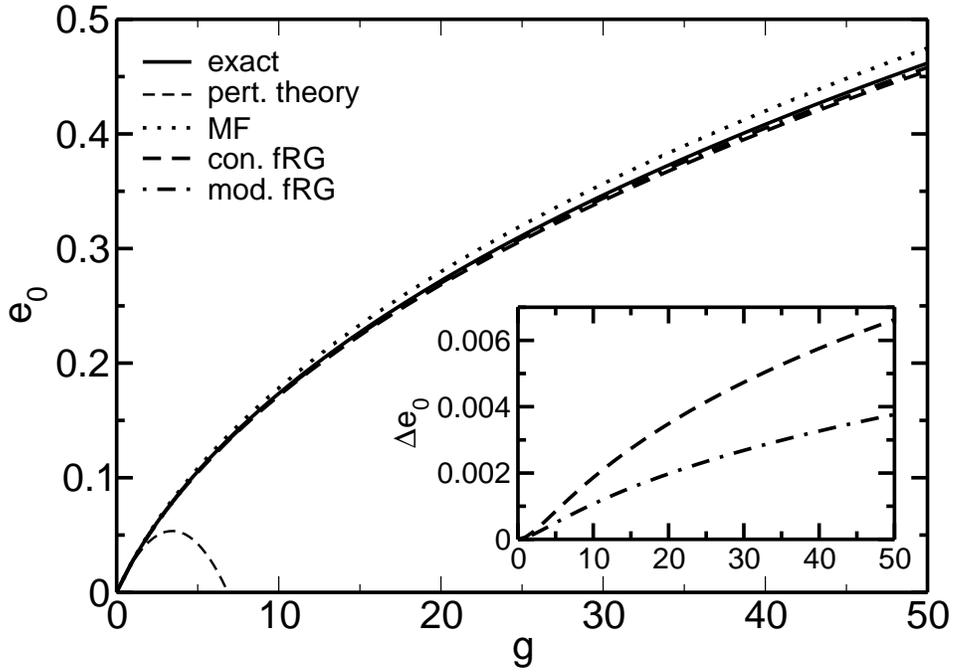}}
\caption[]{Coupling constant dependent part of the
  groundstate energy $e_0=E_0 - E_0^0$ as a function of $g$. Different
  approximations (second order perturbation theory [thin dashed line],
  mean field [dotted line], conventional fRG [thick dashed line],
  modified fRG [dashed-dotted line]) are compared to the exact result
  (solid line). The inset shows the difference between the exact
  result and the two fRG approximations. 
\label{fig3}}
\end{figure} 

\begin{figure}[htb]
\centerline{\includegraphics[width=0.8\textwidth,clip]{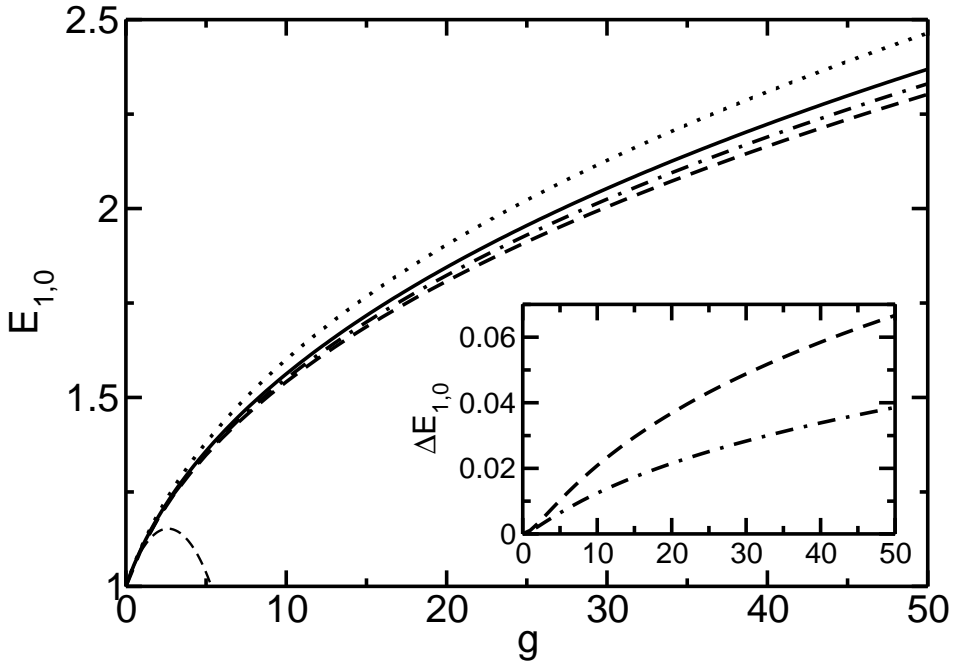}}
\caption[]{As in Fig.\ \ref{fig3}, but for the energy difference
  $E_{1,0}$ of the first excited state and the ground state.
\label{fig4}}
\end{figure} 

\begin{figure}[htb]
\centerline{\includegraphics[width=0.8\textwidth,clip]{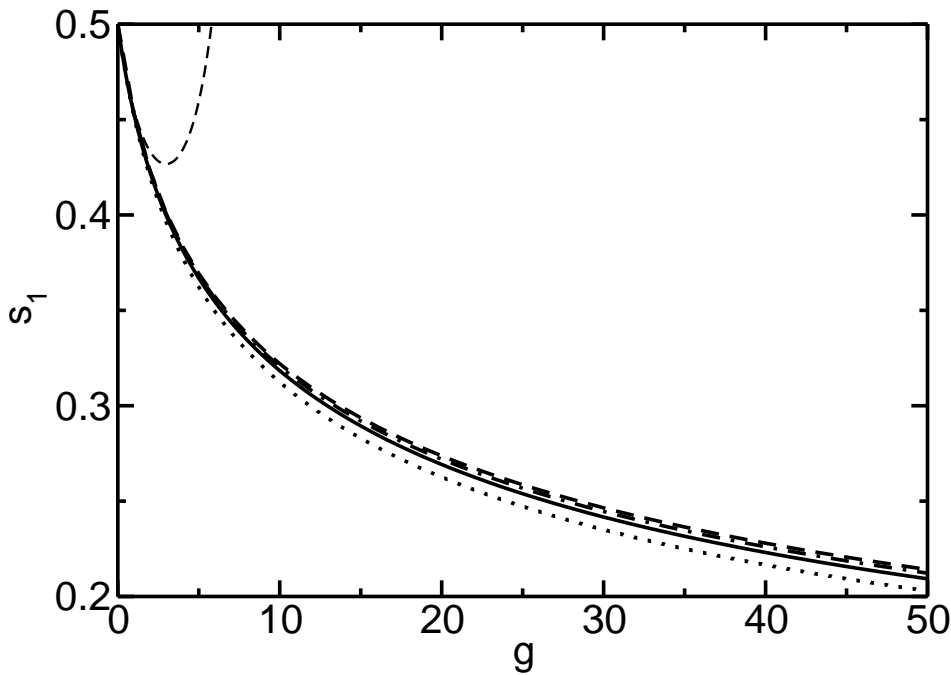}}
\caption[]{As in Fig.\ \ref{fig3}, but for the spectral
  weight $s_1$ of the first peak.
\label{fig5}}
\end{figure}

To numerically solve the set of differential equations (\ref{gamma0ao}), 
(\ref{gamma1ao}), and (\ref{gamma2ao}) we have discretized the
frequencies (which at $T=0$ are continuous) on a linear mesh 
$\omega_j = j \delta$ with $j = -j_0,-j_0+1, \ldots, j_0$
\cite{footnote3}.
By increasing $j_0$ and decreasing $\delta$ convergence can be
achieved up to the required accuracy. For our purposes $j_0=40$ and
$\delta=0.5$ turned out to be appropriate. This leads to a set of 
roughly $5.3\times 10^5$ coupled equations. 
The $\Lambda$-integration is started at $\Lambda_0=10^5$ making sure 
that further increasing $\Lambda_0$ does not lead to significant 
changes in the results. 
Figs.\ \ref{fig3} to \ref{fig5} show comparisons of $e_0$, 
$E_{1,0}$, and $s_1$ with the different approximations considered here 
(second order perturbation theory, mean field, conventional fRG, 
modified fRG) and the exact results. Although the approximate 
fRGs correctly reproduce only the first two derivatives with 
respect to $g$ at $g=0$, they give extremely accurate 
results even up to $g=50$, while conventional second order perturbation
theory can only be trusted for $g<1$. This provides an impressive
example of the power of ``renormalization group enhanced perturbation
theory''. Comparing the two fRG approximations the modified scheme is
roughly a factor of two closer to the exact result and thus
indeed a substantial improvement \cite{katanin,footnote5}.
  
To avoid the problem of analytic continuation (see the next section) 
the results for $e_{1,0}^{\rm fRG}$ and $s_1^{\rm fRG}$ were obtained
by fitting a function $a/(\omega^2 + b^2)$ with $a$ and $b$ as fitting
parameters to ${\mathcal G}^{\rm fRG}(i \omega) = \left[ \omega^2+1 -
\Sigma^{\rm fRG}(i\omega) \right]^{-1}$. 
Assuming this fitting form we have used that the spectral
function is dominated by the first peak \cite{footnote4}.  
For the problem studied also mean field theory leads to fairly
accurate results (but not as good as the fRG).  This 
is related
to the fact that low-lying eigenstates of the Hamiltonian in 
Eq.\ (\ref{anhaos}) can be
described quite well by the eigenstates of a harmonic oscillator with a 
shifted frequency determined self-consistently.

\section{Single-particle dynamics of the single impurity Anderson model}
\label{sec:SIAM}
In contrast to the anharmonic oscillator studied in the previous section,
the single impurity Anderson model \cite{SIAM}
\begin{equation}\label{equ:SIAM}
H=\sum\limits_{\vec k\sigma}\epsilon_{\vec k}
c^\dagger_{\vec k\sigma}c^{\phantom{\dagger}}_{\vec k\sigma}
+
\epsilon_d\sum\limits_\sigma d^\dagger_{\sigma}d^{\phantom{\dagger}}_{\sigma}
+U d^\dagger_{\uparrow}d^{\phantom{\dagger}}_{\uparrow}
d^\dagger_{\downarrow}d^{\phantom{\dagger}}_{\downarrow}
+
\frac{V}{\sqrt{N}}\sum\limits_{\vec k\sigma}\left(
c^\dagger_{\vec k\sigma}d^{\phantom{\dagger}}_{\sigma}+\mbox{h.c.}\right)
\end{equation}
consists of two subsystems, namely a ``conduction band'' with continuous energy
spectrum described by the first term in Eq.\ (\ref{equ:SIAM}) and
a localized level
(typically referred to as ``$d$''-state) with energy $\epsilon_d$ relative to the
chemical potential $\mu$ of the conduction electrons. Two electrons
occupying the localized level are
in addition
subject to a Coulomb repulsion $U>0$.
Both subsystems are coupled via the hybridization in the last term. As
the two-body interaction is restricted to a single level the SIAM falls into
the of class of zero-dimensional interacting systems.

While this model looks rather simple, it contains all ingredients that
make it a complicated many-body problem. The bare energy scales of the model
are the bandwidth $W$ of the band electrons, the local energy $\epsilon_d$, the
Coulomb repulsion $U$, and the bare level width generated by
the hybridization, $\Delta_0=\pi V^2\rho_c(\mu)$, where
$\rho_c(\epsilon)$
 denotes
the density of states of the conduction states which is assumed to be 
slowly varying.
In the following we restrict ourselves to the
particle-hole symmetric case $2\epsilon_d+U=2\mu=0$.
As in the parameter regime $U/\Delta_0\gg1$ the charge fluctuations on
the $d$-level from the average value $1$ are small, 
 it  can  effectively be described by a spin antiferromagnetically
coupled to the conduction electron spin density \cite{swt}, the so-called
Kondo model \cite{Kondo}. This antiferromagnetic coupling leads to a screening
of the local spin by the conduction electrons with a characteristic energy
scale $\ln(T_{\rm K})\propto-1/\Delta_0$, the Kondo temperature.
As is apparent from this expression, $T_{\rm K}$ depends non-analytically on
$\Delta_0$, signalling the occurrence of infrared divergences in perturbation
theories in $ \Delta_0/U$ 
\cite{hewson} and a severe problem for computational techniques,
namely the task to resolve an exponentially small energy scale.

On the other hand, the physics in the regimes $T,\omega\gg T_{\rm K}$ and
$T,\omega\ll T_{\rm K}$ is comparatively simple. For $T,\omega\gg T_{\rm K}$
it is governed by charge excitations with energies $\epsilon_d$ and
$\epsilon_d+U$ with a life-time given by $1/(2\Delta_0)$.
In the other limit $T,\omega\ll T_{\rm K}$ it has been worked out using
Wilson's NRG \cite{Wilson} that the system
can again be described by a Hamiltonian of the form Eq.\ (\ref{equ:SIAM}), but with
$U\to U^\ast=0$, $\epsilon_d\to\epsilon_d^\ast=0$ and
$\Delta_0$ replaced by $\Delta_0^\ast\sim T_{\rm K}$. This regime has been
coined ``local Fermi liquid'' by Nozi\`eres \cite{hewson}. This
effective description yields a narrow resonance at the chemical
potential in the spectral
function of the $d$-Green function which in the symmetric case takes the form 
\begin{equation}
G_{dd}(i\nu)=\frac{1}{i\nu-\Delta (i\nu)-\tilde \Sigma(i\nu)}.
\end{equation}
Here $\Delta(i\nu)=V^2\int d\epsilon \rho_c(\epsilon)/(i\nu-\epsilon)$
and  $\tilde \Sigma$ denotes all self-energy contributions of second and higher order in $U$. As a special feature of the symmetric case, already the
approximation to only keep the 
second order self-energy $\Sigma^{(2)}$ describes the qualitative
behavior of the $d$-spectral function for different values of
$U/\Delta_0$ correctly \cite{hewson}. For $U/\Delta_0 \ll 1$ there is a Lorentzian peak
at the chemical potential with a width differing little from
$\Delta_0$. For $U/\Delta_0 \gg 1$ most of the spectral weight is in
the high energy peaks near $\pm U/2$ with a narrow resonance at
$\mu=0$. Quantitatively the width and shape of this Kondo (or
Abrikosov-Suhl) resonance is described
poorly using $\Sigma^{(2)}$,  
vanishing only $\sim \Delta_0/U$. From the results of Sect.\ 3
one expects that the application of the
functional
 renormalization group to the SIAM  leads to  improvements over the
direct perturbational  
description of the Kondo resonance. For the SIAM the fRG results can
be compared to the outcome of NRG calculations
\cite{RG_general,Wilson}.  An important
aspect of the treatment of the SIAM with the fRG is that, while the 
numerical effort
of the NRG method increases exponentially with the number of impurity degrees
of freedom, e.g.\ in case of additional orbital degrees of freedom or for 
a system of many magnetic impurities,
the increase in computational resources necessary in the fRG approach described
below is governed at most by a power-law.
 
The starting point of our fRG approach are again Eqs. (\ref{gamma1fl})
and (\ref{gamma2fl}) . We also use $m_c=2$, i.e. we replace the
3-particle-vertex by its initial condition
$\gamma_3^\Lambda=0$,
and consider $T=0$ only.
We again use the frequency cutoff described in Eq.\ (\ref{matsubaracutoff})
and because we presently concentrate on spectral properties
only flow equations for the self-energy and for the 2-particle-vertex 
(and not the ground state energy) will
be derived. Note that the calculation of two-particle properties is possible
within the same scheme without further difficulties \cite{metzner}
and will be discussed in a forthcoming publication.

For the SIAM all Green functions $G_{i,j}$ different from $G_{dd}$
can be expressed in terms of $G_{dd}$ and Green functions of the
non-interacting system. For the calculation of  $G_{dd}$
 the multi-index $k$ is replaced by a spin-index $\sigma$ and a
 frequency $\nu$. The 2-particle-vertex can then be written as
\begin{eqnarray}\label{gamma2:SIAMa}
\hspace{-2cm}
\g(k_1',k_2';k_1,k_2)=\D{\nu_1+\nu_2-\nu_1'-\nu_2'}\\\nonumber
\times\Bigl\{\DS \delta_{\sigma_1,\sigma_1'}\delta_{\sigma_2,\sigma_2'}\ \gw(i\nu_1',i\nu_2';i\nu_1,i\nu_2)
-\DS \delta_{\sigma_1,\sigma_2'}\delta_{\sigma_2,\sigma_1'}\ \gw(i\nu_2',i\nu_1';i\nu_1,i\nu_2)
\Bigr\}.\end{eqnarray}
This can be used to perform the sum over the spins in
Eq.\ (\ref{gamma1fl})
 and we find for the self-energy

\begin{equation}\label{Self:SIAM}
\hspace{-2cm}
\DDL\Se{i\nu}=-\frac{1}{2\pi}\int_{-\infty}^{\infty}d\nu'
\Gp{i\nu'}\Bigl[
2 \, \gw(i\nu,i\nu';i\nu,i\nu')-
\gw(i\nu',i\nu;i\nu,i\nu')
\Bigr].
\end{equation}
In order to derive a flow equation for
$\gw(i\nu_1',i\nu_2';i\nu_1,i\nu_2)$, with
$\nu_2=\nu_1'+\nu_2'-\nu_1$, the spin-sum in Eq.\ (\ref{gamma2fl})
has to be performed as well. This leads to the lengthy 
expression Eq.\ (\ref{gamma2:SIAM}), presented in Appendix A. 

The initial values are given by
$\Sigma^{\Lambda=\infty}(i\nu)=0$ and $\mathcal U^{\Lambda=\infty}(i\nu_1',i\nu_2';i\nu_1,i\nu_2)=U$.
In the numerical solution of these flow equations we start integrating
at finite $\Lambda_0 \gg \max (U,\Delta_0)$. The integration from $\Lambda=\infty$ to $\Lambda_0$ can be performed analytically. Up to corrections of order $\Lambda_0^{-1}$ the new initial conditions are $\Sigma^{\Lambda_0}(i\nu)=U/2$ and $\mathcal U^{\Lambda_0}(i\nu_1',i\nu_2';i\nu_1,i\nu_2)=U$.

$\mathcal S^\Lambda(i\nu)$ and $\mathcal P_{\rm con}^{\Lambda}(i\nu,i\nu')=\mathcal S^\Lambda(i\nu) \mathcal G^\Lambda(i\nu')$ are calculated the same
way as for the anharmonic oscillator using Eq.\ (\ref{morristrick}),
 and are given by
\begin{equation}
  \label{S:SIAM}
\Gp{i\nu} \to
\D{|\nu|-\Lambda}\frac{1}{\left[\Gn{i\nu}\right]^{-1}-\Se{i\nu}} \; ,
\end{equation}
with $\left[\Gn{i\nu}\right]^{-1} = i \nu - \Delta(i \nu) - \epsilon_d$
and
\begin{equation}
\hspace{-1cm}
\mathcal P_{\rm con}^{\Lambda}(i\nu,i\nu') \to\D{|\nu|-\Lambda}\frac{1}{\left[\Gn{i\nu}\right]^{-1}-\Se{i\nu}}
\frac{\H{\nu'}}{\left[\Gn{i\nu'}\right]^{-1}-\Se{i\nu'}}
\end{equation}
with $\Theta(0)=1/2$.
In our calculations we use the limit of an infinite bandwidth, i.e. 
$\Delta(i\nu)\to -i \, {\rm sign} (\nu) \, \Delta_0$.
 The results of second
order perturbation theory for the self-energy
  can be recovered by replacing the self-energy on the rhs of
Eqs.\ (\ref{Self:SIAM}) and (\ref{gamma2:SIAM}) and $\mathcal
U^\Lambda$
 on the rhs
of Eq.\ (\ref{gamma2:SIAM}) by their initial values.
It turns out that the fRG version 
using $ \mathcal P_{\rm con}^{\Lambda} $
gives good results for
$U/\Delta_0<3$, but  fails to provide the expected improvement compared
to the use of $\Sigma^{(2)}$ for $U/\Delta_0 > 3$. This will
be discussed in more detail in a forthcoming publication.
For this reason we follow \cite{katanin} and replace
 $\mathcal P^\Lambda_{\rm con}(i\nu,i\nu') $ by $\mathcal
 P^\Lambda_{\rm mod}(i\nu,i\nu')=-\mathcal G^\Lambda(i\nu')
\DS\DDL\mathcal G^\Lambda(i\nu)$ in Eq.\ (\ref{gamma2:SIAM}).
Applying Eq.\ (\ref{morristrick}) yields 
\begin{equation}
\label{Kat:SIAM}
\hspace{-2cm}
\begin{array}{lll}
-\Gl{i\nu'}\DS\DDL\Gl{i\nu}
&\to&\DS\D{|\nu|-\Lambda}\frac{1}{\left[\Gn{i\nu}\right]^{-1}-\Se{i\nu}}
\frac{\H{\nu'}}{\left[\Gn{i\nu'}\right]^{-1}-\Se{i\nu'}}\\[8mm]
&-&\DS\DDL\Se{i\nu}~\frac{\H{\nu}}{[\left[\Gn{i\nu}\right]^{-1}-\Se{i\nu}]^2}
\frac{\H{\nu'}}{\left[\Gn{i\nu'}\right]^{-1}-\Se{i\nu'}}.
\end{array}
\end{equation}

For the integration of the $T=0$-flow equations the continuous frequencies
again  have to be discretized. However, in contrast to
the anharmonic oscillator it is not sufficient to work with
 a linear mesh. Because for $U/\Delta_0 \gg 1$ the Kondo resonance 
is visible only on an
exponentially small energy scale around the Fermi level,
 we use a combination of a linear and a logarithmic mesh.
This enables us to recover both, the high energy 
 as well as the low-energy physics keeping the number of frequencies to a manageable
size.
 The numerical effort grows with the third power of the number of
 frequencies in the conventional version and with almost
the forth power using the modified version,
 due to the absence of a $\delta$-function in the last term of
 Eq.\ (\ref{Kat:SIAM}).

In the fRG one naturally obtains
 the self-energy for imaginary frequencies.
In order to calculate spectral functions, an
analytic continuation to the real axis is necessary, which is known
to be an  ill-posed problem. Since the results from fRG are not subject
to statistical errors or noise as for example quantum Monte Carlo data,
we have applied the method of Pad\'e approximation \cite{Pade} to obtain
$\Sigma(z)$ from $\Sigma(i\nu)$.
 We find that especially the low-energy part
of the spectral function which contains the Kondo resonance can be
reliably
 extracted if sufficiently many frequencies close to the Fermi level
are used (see below).

 \begin{figure}[htb]
\centerline{\includegraphics[width=0.9\textwidth,clip]{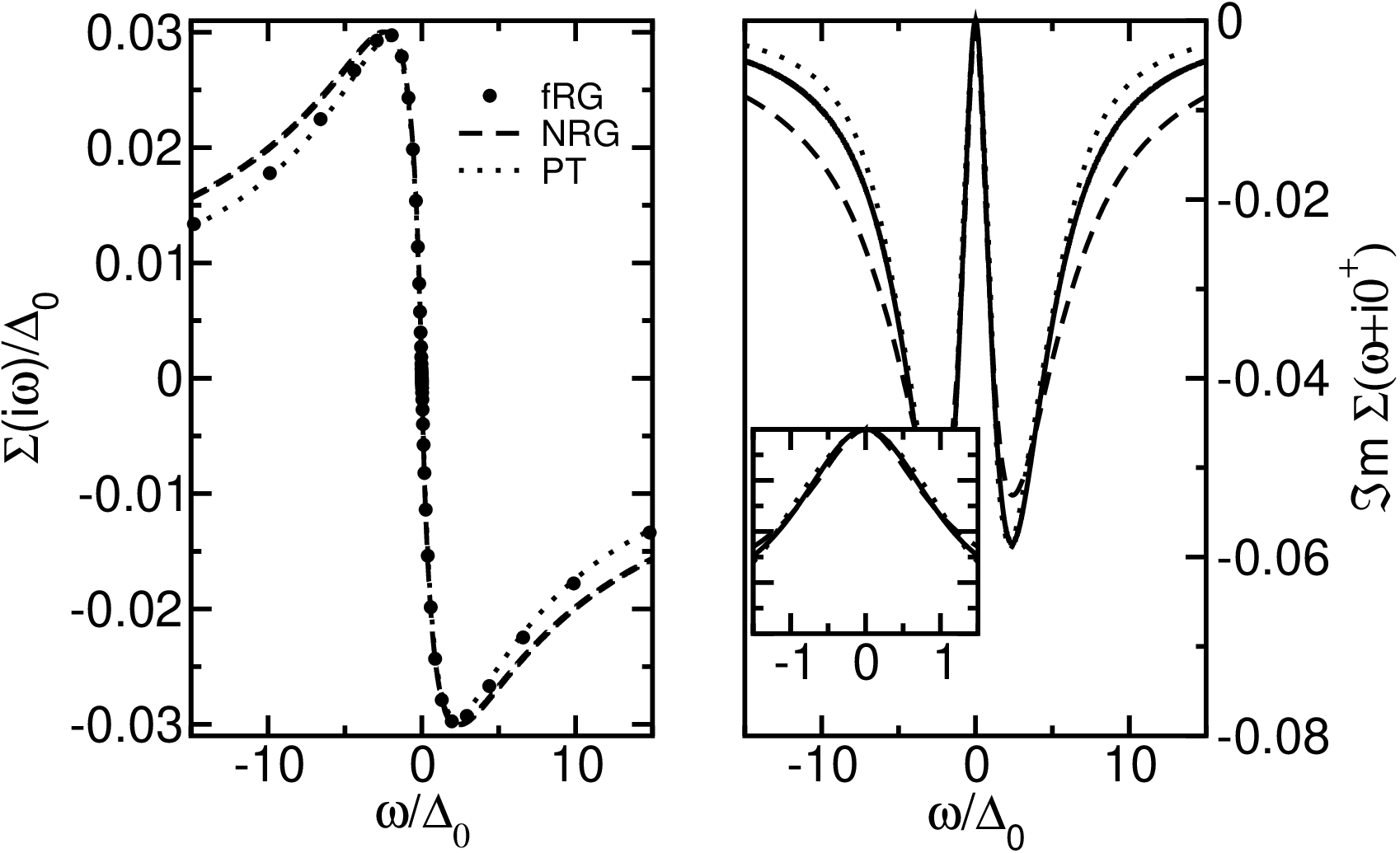}}
\caption[]{Left panel: $\Sigma(i\omega)$ from fRG for $U=\Delta_0$ (circles)
compared to NRG (dashed line) and second order perturbation theory (dotted line).\\
Right panel: $\Im m\Sigma(\omega+i0^+)$ for fRG obtained from Pad\'e approximants
for the data in left panel. The inset shows an enlarged view around $\omega=0$.
\label{fig:SIAM1}}
\end{figure}

As an example, we present in Fig.\ \ref{fig:SIAM1} the calculated $\Sigma(i\omega)$
(left panel) and the corresponding $\Im m\Sigma(\omega+i0^+)$ from the
Pad\'e approximation (right panel) for $U/\Delta_0=1$. The dashed and dotted
lines represent
results from NRG and second order perturbation theory, respectively. The inset in the right panel
shows an enlarged view of the region around $\omega=0$. 
Apparently, the Pad\'e approximation
to the fRG provides reliable results, in particular around $\omega=0$, and
recovers the perturbation theory as expected for such small value of $U$.
Discrepancies to NRG for large $\omega$ can be traced to broadening effects in the NRG.

\begin{figure}[htb]
\centerline{\includegraphics[width=0.9\textwidth,clip]{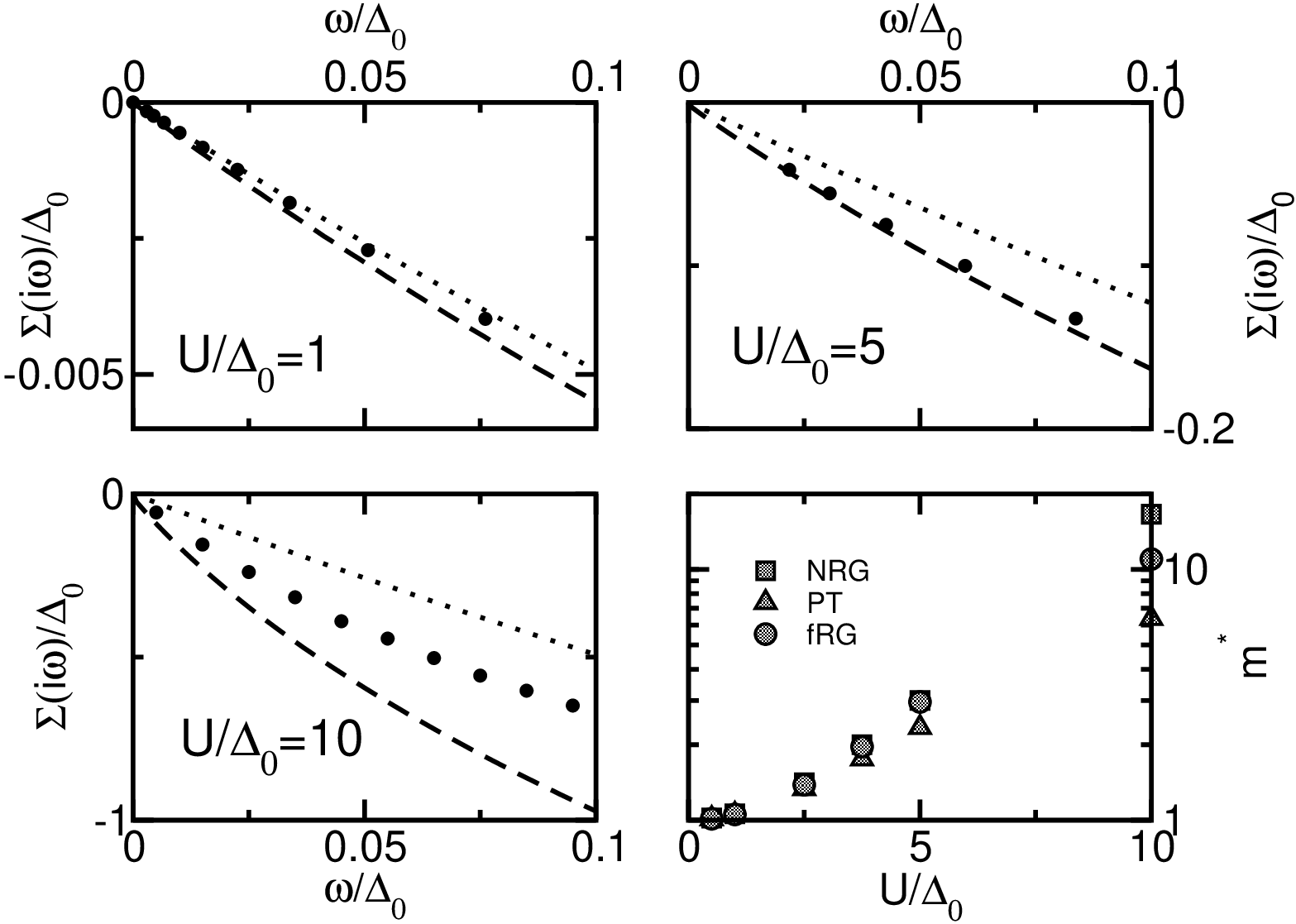}}
\caption[]{$\Sigma(i\omega)$ from fRG (circles)
for $U/\Delta_0=1$, $5$ and $10$ compared to NRG (dashed line) and second order 
perturbation theory (dotted line).\\
Bottom right: Effective mass $m^\ast$ from fRG (circles), NRG (squares) and
second order perturbation theory (triangles).\label{fig:SIAM2}}
\end{figure}

The true test for the method however is a comparison of the
self-energy to NRG results for larger values
of $U$ and in particular the behavior on low-energy scales. 
Such a comparison is presented in Fig.\ \ref{fig:SIAM2} for 
$U/\Delta_0=1$, $5$, and $10$. 
For $U/\Delta_0=5$ the fRG results are still in excellent agreement
with the NRG data while second order perturbation theory already
deviates. For $U=10\Delta_0$ we observe a significant dependency
of the general structures and also of the slope at $\omega\to0$
on details of the discretization mesh. It seems necessary to
have an extremely fine resolution around $\omega\to0$ and a sufficient
resolution around $\omega\approx U/2$. With an exponentially vanishing 
low-energy scale these constraints are hard to fulfill with both
linear and logarithmic
meshes keeping the number of flow equations to a manageable size.
The bottom left pannel contains the best fRG data we were able to 
obtain so far. A more extended discussion of this issue will be 
presented in a forthcoming publication. 

In addition to the self-energy the
effective mass $m^\ast$ obtained from
$$
m^\ast=1+\lim_{\omega\searrow0}\frac{\Im m\Sigma(i\omega)}{\omega}
$$
as a function of $U$ is depicted, in the bottom right panel of
Fig.\  \ref{fig:SIAM2}.
This quantity is directly related to the Kondo temperature \cite{hewson}.
Compared to perturbation theory the effective
mass is much closer to the very accurate values determined from the NRG. 

\begin{figure}[htb]
\centerline{\includegraphics[width=0.9\textwidth,clip]{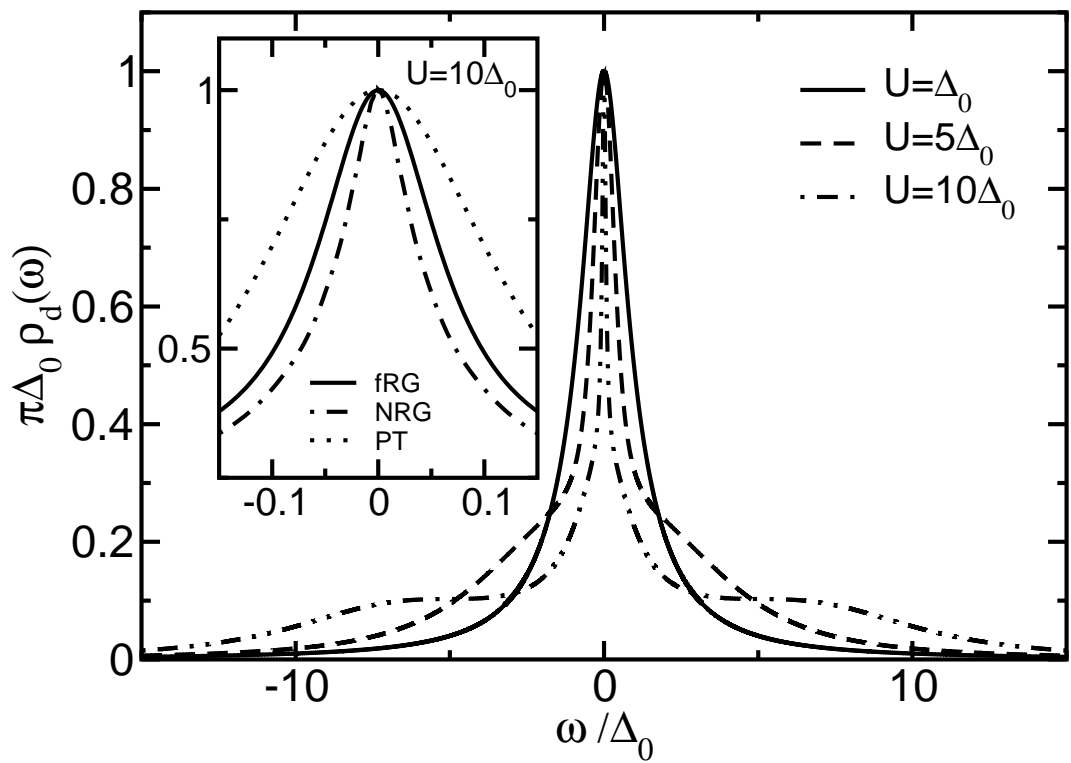}}
\caption[]{The evolution of the fRG spectral function for the $d$-level
for the values of $U$ in Fig.\ \ref{fig:SIAM2}. The inset shows a comparison to NRG for $U/\Delta_0=10$ for the region around 
$\omega=0$.\label{fig:SIAM3}}
\end{figure}

The evolution of the spectral function of the $d$-level,
$\rho_d(\omega)=-\frac{1}{\pi}\Im mG_d(\omega+i 0^+)$ for the above
values of $U$ is collected in
Fig.\ \ref{fig:SIAM3}. The inset shows a comparison to 
NRG and perturbation theory for the region around 
$\omega=0$ at $U/\Delta_0=10$. One can see very nicely the development of the
sharp resonance in the spectrum and the formation of the Hubbard bands
with increasing $U$.
Due to the insufficient resolution at larger $\omega$ in the calculation
for $U/\Delta_0=10$, the Hubbard bands come
out too broad here. We believe that an improved discretization of the 
energy mesh in the solution of the flow equations will remedy that 
particular problem.
The important region around $\omega=0$ on the other hand is captured rather
well by the fRG. It is in particular noteworthy that the functional form
at low  frequencies (see the inset) apparently does not follow a Lorentzian but rather,
like the NRG, the more complex scaling form with logarithmic tails predicted
by Logan {\em et al.} \cite{logan}. 
%
%

\section{Summary and outlook}
\label{sec:summary}
We have presented an application of the functional renormalization
group technique
to solve for the dynamics of zero-dimensional interacting quantum problems.
As particular examples, we discussed the anharmonic oscillator and the
single impurity Anderson model. In both cases the fRG proved to be a
substantial improvement over conventional low-order perturbation theory and
rather close to the very accurate results obtained numerically. 

We also investigated the differences between conventional fRG and a modification
suggested by Katanin \cite{katanin}, which should improve the accuracy
of the method further. That this is indeed true was shown directly for the
anharmonic oscillator. For the single impurity Anderson model it was
actually necessary to use this modified version to obtain sensible results
for values $U/\Delta_0>3$.

An important aspect of the fRG is that an extension of the SIAM to
more complex systems,
like e.g.\ orbital degrees of freedom or systems of coupled impurities
is straightforward. This is in principle also true for Wilson's NRG. 
In the latter, however, the exponentially increasing Hilbert
space renders a practical application quickly impossible. For the fRG, on the
other hand, the major modification will be an increase of the number
of equations, which means that the numerical effort will increase at most
with a power-law. This feature makes the fRG a possible method to study
features of complex impurity systems and in particular a potential
``impurity solver'' for mean-field theories of interacting lattice models
like the Hubbard model in the framework of the dynamical mean-field theory
\cite{rmp_geo} or the dynamical cluster approximation \cite{rmp_tm}.
Furthermore, as it becomes clear from the anharmonic oscillator, the fRG is
of equal complexity for bosonic and fermionic systems. This makes opens
the possibility to study combinations of such degrees of freedom 
in impurity models.
A question which can be addressed using the fRG is the influence of phonons or
magnetic fluctuations on low energy scales. Within the dynamical
mean-field theory the metal-insulator transition
in the presence of phonons (Kondo volume collapse \cite{allen}) as well
as the problem of non-Fermi liquid formation in for example 
CeCu$_{6-x}$Au$_x$ \cite{qimiao} can be studied.

\ack
We thank S.\ Dusuel, W.\ Metzner, M.\ Salmhofer, H.\ Schoeller,
and G.\ Uhrig for useful discussions. This work was supported by
the SFB 602 of the Deutsche Forschungsgemeinschaft (R.H.\ and K.S.).
V.M.\ is grateful to the Bundesministerium f\"ur Bildung und Forschung
for support.

\appendix
\section{}
Depending on the specific parameterization of the
 2-particle-vertex flow equations
for the relevant frequency dependent parts of the vertex and for the self-energy can be derived
by using the spin conservation
on the vertex when the sum over spins in Eq.\ (\ref{gamma2fl})
is performed. For the form of the the
vertex given in Eq.\ (\ref{gamma2:SIAMa}) the corresponding flow equation for the self-energy is given by
Eq.\ (\ref{Self:SIAM}), and the equation for 
the frequency dependent part of the vertex
$\mathcal U^\Lambda(i\nu_1',i\nu_2';\nu_1,i\nu_2)$ with $\nu_2=\nu_1'+\nu_2'-\nu_1$ reads
\begin{eqnarray}
  \hspace{-2cm}
\label{gamma2:SIAM} 
  \DDL \gw\begin{array}[t]{l}\!\!\!\DS(i\nu_1',i\nu_2';i\nu_1,i\nu_2)
=-\frac{1}{2\pi}\int_{-\infty}^{\infty}d\nu
\DS\Biggl[\mathcal P^{\Lambda}(i\nu,i\nu_1+i\nu_2-i\nu)
\\[5mm]
\Bigl(-\gw(i\nu,i\nu_1+i\nu_2-i\nu;i\nu_1,i\nu_2)
\gw(i\nu_2',i\nu_1';i\nu_1+i\nu_2-i\nu,i\nu)\\[5mm]
-\gw(i\nu_1+i\nu_2-i\nu,i\nu;i\nu_1,i\nu_2)
\gw(i\nu_1',i\nu_2';i\nu_1+i\nu_2-i\nu,i\nu)\Bigr)\\[5mm]
\DS+\biggl\{\mathcal P^\Lambda(i\nu,-i\nu_1+i\nu_1'+i\nu)
\\[5mm]
\Bigl(2\, \gw(i\nu_1',i\nu;i\nu_1,-i\nu_1+i\nu_1'+i\nu)
\gw(i\nu_2',-i\nu_1+i\nu_1'+i\nu;i\nu_2,i\nu)\\[5mm]
-
\gw(i\nu_1',i\nu;i\nu_1,-i\nu_1+i\nu_1'+i\nu)
\gw(-i\nu_1+i\nu_1'+i\nu,i\nu_2';i\nu_2,i\nu)\\[5mm]
-
\gw(i\nu,i\nu_1';i\nu_1,-i\nu_1+i\nu_1'+i\nu)
\gw(i\nu_2',-i\nu_1+i\nu_1'+i\nu;i\nu_2,i\nu)\Bigr)\\[5mm]
+(1'\leftrightarrow2'; 1\leftrightarrow2)\biggr\}\\[5mm]
\DS-\biggl\{\mathcal P^\Lambda(i\nu,-i\nu_1+i\nu_2'+i\nu)
\\[5mm]
\gw(i\nu,i\nu_2';i\nu_1,-i\nu_1+i\nu_2'+i\nu)
\gw(-i\nu_1+i\nu_2'+i\nu,i\nu_1';i\nu_2,i\nu)\\[5mm]
+(1'\leftrightarrow2'; 1\leftrightarrow2)\biggr\}
\Biggr] \; .  \end{array} 
\end{eqnarray}
Again $\mathcal P$ stands either for ${\mathcal P}_{\rm con}$ or
${\mathcal P}_{\rm mod}$. 

Two other possible parameterizations of the 2-particle vertex are
\begin{eqnarray*}
\hspace{-2cm}
\g(\xi_1',\xi_2';\xi_1,\xi_2)&=&
\begin{array}[t]{l}\D{\nu_1+\nu_2-\nu_1'-\nu_2'}
\DS \Bigl\{\delta_{\sigma_1,\sigma_2}\delta_{\sigma_1',\sigma_2'}\delta_{\sigma_1,\sigma_1'}\ \gp(i\nu_1',i\nu_2';i\nu_1,i\nu_2)\\[5mm]
\hspace{-2cm} +\DS \delta_{\sigma_1,-\sigma_2}\delta_{\sigma_1',-\sigma_2'}\ \Bigl(\delta_{\sigma_1,\sigma_1'}\ga(i\nu_1',i\nu_2';i\nu_1,i\nu_2)-
\delta_{\sigma_1,-\sigma_1'}\ga(i\nu_2',i\nu_1';i\nu_1,i\nu_2)\Bigr)
\Bigr\}.\end{array}
\end{eqnarray*}\\[5mm]
and
\begin{equation*}
\hspace{-2cm}
\g(\xi_1',\xi_2';\xi_1,\xi_2)=\D{\nu_1+\nu_2-\nu_1'-\nu_2'}\Bigl\{
\begin{array}[t]{l}\DS S_{\sigma_1', \sigma_2'; \sigma_1, \sigma_2}\ \gs(i\nu_1',i\nu_2';i\nu_1,i\nu_2)\\[5mm]
+\DS T_{\sigma_1', \sigma_2'; \sigma_1, \sigma_2}\ \gt(i\nu_1',i\nu_2';i\nu_1,i\nu_2)\Bigr\}\end{array}
\end{equation*}
with
$$
S_{\sigma_1', \sigma_2'; \sigma_1, \sigma_2}=\frac{1}{2}\left(\delta_{\sigma_1, \sigma_1'}\delta_{\sigma_2, \sigma_2'}-\delta_{\sigma_1, \sigma_2'}\delta_{\sigma_2, \sigma_2'}\right);\
T_{\sigma_1', \sigma_2'; \sigma_1,
  \sigma_2}=\frac{1}{2}\left(\delta_{\sigma_1,
    \sigma_1'}\delta_{\sigma_2, \sigma_2'}+\delta_{\sigma_1,
    \sigma_2'}\delta_{\sigma_2, \sigma_2'}\right) \; .
$$
Both these parameterizations lead to different set of
flow equations for the self-energy and to two sets of flow equations
for the frequency dependent functions $\gp$ and $\ga$ for the first
and $\gs$ and $\gt$ for the second (compared to one set for the
parameterization Eq.\ (\ref{gamma2:SIAMa})) implying an increased numerical
effort. They are usefull to further investigate the processes
occurring during the integration of the flow equations.  This
way we are able to distinguish the behavior
of different channels of the interaction.

\setcounter{section}{1}

\end{document}